\documentclass[twocolumn,american,prl,superscriptaddress]{revtex4}
\usepackage[T1]{fontenc}
\usepackage[latin1]{inputenc}
\usepackage{graphicx}
\usepackage{amssymb}

\makeatletter

\usepackage{ae,aecompl}

\usepackage{babel}
\makeatother
\begin{document}

\title{Vortex transmutation}

\author{Albert Ferrando}
\homepage[Interdisciplinary Modeling Group (InterTech): ]{http://www.upv.es/intertech}
\affiliation{Departament d'Òptica, Universitat de València. Dr. Moliner, 50. E-46100
Burjassot (València), Spain.}

\author{Mario Zacarés}
\homepage[Interdisciplinary Modeling Group (InterTech): ]{http://www.upv.es/intertech}
\affiliation{Departament d'Òptica, Universitat de València. Dr. Moliner, 50. E-46100
Burjassot (València), Spain.}

\author{Miguel-Ángel García-March}
\homepage[Interdisciplinary Modeling Group (InterTech): ]{http://www.upv.es/intertech}
\affiliation{Departamento de Matemática Aplicada, Universidad Politécnica de Valencia.
Camino de Vera, s/n. E-46022 Valencia, Spain.}

\author{Juan A. Monsoriu}
\homepage[Interdisciplinary Modeling Group (InterTech): ]{http://www.upv.es/intertech}
\affiliation{Departamento de Física Aplicada, Universidad Politécnica de Valencia.
Camino de Vera, s/n. E-46022 Valencia, Spain.}

\date{\today}

\begin{abstract}
Using group theory arguments and numerical simulations, we demonstrate
the possibility of changing the vorticity or topological charge of
an individual vortex by means of the action of a system possessing
a discrete rotational symmetry of finite order. We establish on theoretical
grounds a {}``transmutation pass rule'' determining the conditions
for this phenomenon to occur and numerically analize it in the context
of two-dimensional optical lattices or, equivalently, in that of Bose-Einstein
condensates in periodic potentials. 
\end{abstract}


\maketitle

Vortices are a physical phenomenon common to all complex waves.
Defined by a phase singularity implying the vanishing of the wave
amplitude their presence is ubiquitous in physics where examples of
vortices can be found in as diverse systems as quantized superfluids
and superconductors, Bose-Einstein condensates (BEC's), nonlinear
optical structures, or low dimensional condensed matter or particle
systems (for a review see \cite{desyatnikov-arXiv:nlin_0501026,rebbi85}).
The possibility of changing vortex properties using periodic systems
is a natural step based on the known example of the different behavior
of electrons with or without the presence of a crystal. Like electrons,
properties of vortices in a lattice have been shown to be qualitatively
different than in a homogeneous medium. Vortices have been numerically
predicted to exist in two-dimensional (2D) arrays of coupled waveguides
\cite{malomed-pre64_026601}, in 2D periodic dielectric media with
Kerr nonlinearities ---and, equivalently, in 2D BEC's with periodic
potentials--- \cite{yang-ol28_2094,baizakov-eurol63_642} and in photonic
crystal fibers with defects \cite{ferrando-oe12_817}. They have been
experimentally observed in optically-induced square photonic lattices
\cite{neshev-prl92_123903,fleischer-prl92_123904}. In all cases the
presence of the periodic medium has a strong influence on vortex features
thus opening a door for their external manipulation. In this Letter
we will show how manipulation by means of an external system owning
discrete rotational symmetry can even affect the most intrinsic feature
of a vortex, its vorticity or topological charge, leading to a phenomenon
that we refer to as {}``vortex transmutation''. 

Angular momentum is conserved in a nonlinear medium with $O(2)$ rotational
symmetry in the $x$-$y$ plane described by a first-order evolution
equation of the type $L(|\phi|)\phi=-i\partial\phi/\partial z$ for
the complex scalar field $\phi$. If we consider a solution with well-defined
angular momentum $l\in\mathbb{Z}$ (i.e., an eigenfunction of the
angular momentum operator $-i\partial/\partial\theta$: $\phi_{l}=e^{il\theta}f_{l}(r)$)
at a given axial point $z_{0}$, evolution will preserve the value
of $l$ for all $z$. In a system possessing a discrete point-symmetry
(described by the \emph{$\mathcal{C}_{n}$} and $\mathcal{C}_{nv}$
groups) angular momentum is no longer conserved. However, in this
case one can define another quantity $m\in\mathbb{Z}$, the Bloch
or pseudo-angular momentum, which is conserved during propagation
\cite{ferrando-arXiv:nlin_0411059}. The pseudo-angular momentum $m$
plays then the role of $l$ in a system with discrete rotational symmetry.
From the group theory point of view, the angular and pseudo-angular
momenta $l$ and $m$ are also the indices of the 2D irreducible representations
of $O(2)$ and $\mathcal{C}_{n}$, respectively \cite{ferrando-oe13_1072,ferrando-arXiv:nlin_0411005,hamermesh64}.
Unlike $l$, the values of $m$ are limited by the order of the point-symmetry
group $\mathcal{C}_{n}$: $|m|\le n/2$ \cite{ferrando-arXiv:nlin_0411005,ferrando-arXiv:nlin_0411059}.

\begin{figure}
\includegraphics[%
  scale=0.3]{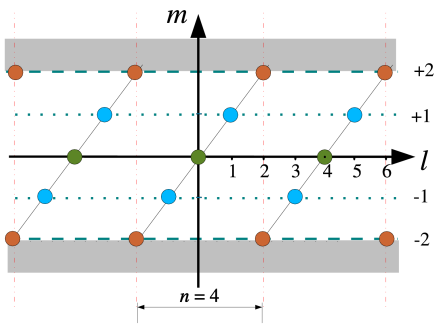}

\caption{Allowed values of the pseudo-angular momentum $m$ in a $\mathcal{C}_{4}$
medium in terms of the angular momentum $l$ of the field $\phi_{l}$
impinging the interface from an $O(2)$ medium.\label{cap:allowed_angular_momentum}}
\end{figure}

The appearance of this upper bound for the pseudo-angular momentum
$m$ opens the interesting question of determining the behavior of
solutions propagating in a $O(2)$ rotational invariant medium with
well-defined angular momentum $l$ after impinging a medium with discrete
symmetry of finite order in which the value of $l$ exceeds the upper
bound for pseudo-angular momentum. This question can be analyzed in
the light of group theory. Let us consider a wave propagating in a
$O(2)$ nonlinear medium corresponding to a solution $\phi_{l}$
(not necessarily stationary) with well-defined angular momentum $l$
launched into a second nonlinear medium characterized by the $\mathcal{C}_{n}$
group. The surface separating the two media defines an $O(2)$-$\mathcal{C}_{n}$
interface that we locate at $z=0$. We assume evolution is first order
in $z$: $L(|\phi|)\phi=-i\partial\phi/\partial z$. Evolution in
the second medium is thus fully determined by the initial condition
$\phi_{l}(0)$. The initial field $\phi_{l}(0)$ will excite a different
representation of the $\mathcal{C}_{n}$ group depending on the value
of $l$. Once this second wave is excited, it will propagate in the
$\mathcal{C}_{n}$ medium by preserving its representation ---defined
by its pseudo-angular momentum $m$. Let $\varphi_{m}$ be a function
in the representation of $\mathcal{C}_{n}$ characterized by the index
$m$. Let us determine now what values of $l$ are allowed by symmetry
to produce a nonzero projection of $\phi_{l}(0)$ onto $\varphi_{m}$
for a given value of $m$. The projection coefficient is given by
$c_{ml}=\int_{\mathbb{R}^{2}}\varphi_{m}^{*}(r,\theta)\phi_{l}(r,\theta;0)$.
Since $\varphi_{m}$ and $\phi_{l}$ belong to representations of
$\mathcal{C}_{n}$ and $O(2)$, respectively, they both properly transform
under a discrete rotation of order $n$: $\varphi_{m}(r,\theta+2\pi/n)=e^{im(2\pi/n)}\varphi_{m}(r,\theta)$
and $\phi_{l}(r,\theta+2\pi/n)=e^{il(2\pi/n)}\phi_{l}(r,\theta)$.
Thus, by performing the change of variable $\theta\rightarrow\theta+2\pi/n$
in the definition of $c_{ml}$ one arrives to the symmetry relation
$c_{ml}=\exp[i(l-m)2\pi/n]c_{ml}$. The $c_{ml}$ coefficient is then
zero unless the following condition is fulfilled:\begin{equation}
l-m=kn\,\,\,(k\in\mathbb{Z}),\,\,\mathrm{where\,\,\,\,\,}|m|\le\frac{n}{2}.\label{eq:angular_momentum_pass_rule}\end{equation}
The $m$ representation of $\mathcal{C}_{n}$ is thus excited by initial
fields having angular momenta $l=m,m\pm n,m\pm2n,\dots$. In a $O(2)$-$O(2)$
interface each representation of angular momentum $m$ in the second
medium is excited by one, and only one, angular momentum component
$l$ arising from the first medium and verifying $l=m$. Contrarily,
in a $O(2)$-$\mathcal{C}_{n}$ interface the symmetry restriction
(\ref{eq:angular_momentum_pass_rule}) implies that, due to the cutoff
$|m|\leq n/2$ in the $\mathcal{C}_{n}$ medium, there are infinite
angular momenta $l$ that can excite a given representation of index
$m$ in the second medium. This can be clearly seen in Fig.~\ref{cap:allowed_angular_momentum}
in which we represent the permitted values of $m$ for different values
of $l$ for the particular case of a second medium with fourth-fold
symmetry ($n=4)$ according to the angular momentum {}``pass rule''
in Eq.(\ref{eq:angular_momentum_pass_rule}). The effect of the cutoff
($m\leq2$) is apparent in this representation.

Thus, when the incident field carries an angular momentum $l$ that
overcomes the limiting value for pseudo-angular momentum in the second
medium it will excite a wave that will propagate with \emph{different}
constant pseudo-angular momentum $m$ given by the {}``pass rule''
(\ref{eq:angular_momentum_pass_rule}). This result is valid for waves
verifying an equation of the type $L(|\phi|)\phi=-i\partial\phi/\partial z$,
linear or nonlinear, stationary or evolving. A particularly interesting
situation is that in which the incident field is a vortex field of
the $O(2)$ nonlinear medium. This vortex field $\phi_{l}^{\mathrm{v}}$ is
a stationary solution of the evolution equation with well-defined
angular momentum $l\neq0$: $L(|\phi_{l}^{\mathrm{v}}|)\phi_{l}^{\mathrm{v}}=-\mu\phi_{l}^{\mathrm{v}}$.
We consider here individual {}``canonical'' vortices with a single
phase singularity (i.e., with only one point in which $\phi_{l}^{\mathrm{v}}=0$):
$\phi_{l}^{\mathrm{v}}(r,\theta,z)=e^{il\theta}f_{l}^{\mathrm{v}}(r)e^{-i\mu z}$ \cite{molina-terriza-prl87_23902}.
The vorticity or topological charge of such solutions will be given
by the circulation of its phase gradient around the singularity $v=(1/2\pi)\oint\nabla\arg(\phi_{l}^{\mathrm{v}})d\mathbf{r}$,
which equals angular momentum for canonical vortices $v=l$. On the
other hand, the propagating wave $\phi_{m}$ with pseudo-angular
momentum $m$ excited by $\phi_{l}^{\mathrm{v}}$ will evolve in the $\mathcal{C}_{n}$
medium. There are different options for the asymptotic states of $\phi_{m}$
when $z\rightarrow\infty$. One possibility is this wave asymptotically
tends to an stationary solution $\phi_{m}\stackrel{z\rightarrow\infty}{\rightarrow}\phi_{m}^{\mathrm{v}}(r,\theta,z)=e^{im\theta}g_{m}(r,\theta)e^{-i\mu'z}$
in the representation of $\mathcal{C}_{n}$ given by the conserved
pseudo-angular momentum $m$. If $\phi_{m}^{\mathrm{v}}$ has a single phase
singularity then it will have the structure of an individual canonical
discrete-symmetry vortex, its vorticity or topological charge $v'$
being directly given by $m$: $v'=m$ \cite{ferrando-arXiv:nlin_0411005}.
The formation in the $\mathcal{C}_{n}$ medium of an asymptotic stationary
state in the form of a discrete-symmetry vortex is a dynamical issue
that depends on the structural parameters of the second medium as
well as in the characteristics of the input vortex field $\phi_{l}^{\mathrm{v}}$.
If dynamics allows the stabilization of the discrete-symmetry vortex
solution, the $O(2)$-$\mathcal{C}_{n}$ interface will realize the
mapping of a $O(2)$ vortex with charge $v=l$ (exceeding the limiting
value for pseudo-angular momentum) into a $\mathcal{C}_{n}$ vortex
with charge $v'=m\neq0$, such that $v'<v$. The {}``pass rule''
for pseudo-angular momentum (\ref{eq:angular_momentum_pass_rule})
becomes a {}``pass rule'' relating input and output vorticities:\begin{equation}
v-v'=kn\,\,\,(k\in\mathbb{Z}),\label{eq:vorticity_pass_rule}\end{equation}
where $v'$ presents a cutoff in terms of $n$ given by:$\left|v'\right|<n/2\,\,\,(\mathrm{even}\,\, n)$
and $\left|v'\right|\le(n-1)/2\,\,\,(\mathrm{odd}\,\, n)$ \cite{ferrando-arXiv:nlin_0411005}.
Note that $m=n/2$ solutions are not vortices but nodal or dipole-mode
solitons \cite{ferrando-arXiv:nlin_0411005}. We will refer to the
process of mapping an individual vortex into another with different
topological charge as {}``vortex transmutation''.

We will provide now a physical example of system in which the phenomenon
of {}``vortex transmutation'' takes place. It is an optical interface
separating two 2D dielectric media with Kerr nonlinearity, these two
media being a homogeneous medium and a 2D square optical lattice.
This system is equivalent to a 2D BEC in which a periodic potential
is abruptly switched on. They constitute an $O(2)$-$\mathcal{C}_{4}$
interface given by the following equation:\begin{equation}
\left(\nabla_{t}^{2}-V(x,y,z)+\gamma(z)|\phi|^{2}\right)\phi=-i\frac{\partial\phi}{\partial z},\label{eq:nonlinear_eq}\end{equation}
in which $\nabla_{t}$ is the 2D gradient operator and $V(x,y,z)=V_{0}+\theta(z)(V_{1}(x,y)-V_{0})$
where $\theta(z)$ is the step function and $V_{0}$ and $V_{1}(x,y)=V_{1}\left(\cos^{2}(\frac{2\pi}{\Lambda}x)+\cos^{2}(\frac{2\pi}{\Lambda}y)\right)$
($V_{1}$ is the potential strength and $\Lambda$ is the lattice
spatial period) define the refractive index profile of the homogeneous
medium and of the 2D optical lattice: $V_{0}=-(n^{2}-n_{0}^{2})$
and $V_{1}(\mathbf{x})=-(n^{2}(\mathbf{x})-n_{0}^{2})$, $n_{0}$
being a reference refractive index introduced by the slowly-varying
envelope approximation. The nonlinear function $\gamma(z)=\gamma+(1-\gamma)\theta(z)$
permits the nonlinear response of the system to be different in the
two media. All distances appearing in Eq.(\ref{eq:nonlinear_eq})
are normalized and dimensionless ($\mathbf{x}=k_{0}\mathbf{x}'$,
$z=k_{0}z'$). In order to solve the evolution problem in this system
we solve first Eq.(\ref{eq:nonlinear_eq}) for $z<0$, which becomes
an ordinary Nonlinear Schr\"odinger equation (NLSE) for a homogeneous
medium. Since our aim is to evidence the phenomenon of {}``vortex
transmutation'' we are interested in finding canonical vortex solitons
of different charges in the homogeneous $O(2)$ medium: $\phi_{l}^{\mathrm{v}}(\mathbf{x},z)=e^{il\theta}f_{l}^{\mathrm{v}}(r)e^{-i\mu z}$.
This can be done by standard methods. At a given value of $l$, a
family of $O(2)$ vortices are found characterized by their power
$P_{l}=\int_{\mathbb{R}^{2}}|\phi_{l}^{\mathrm{v}}|^{2}$ and their propagation
constant $\mu$, which are related through the relation $P_{l}(\mu)$.
In the case of a Kerr nonlinearity, $\mu$ behaves as a scaling parameter
and $P_{l}$ is $\mu$-independent \cite{desyatnikov-arXiv:nlin_0501026}.
Once the vortex solution $\phi_{l}^{\mathrm{v}}$ is found, it is taken as
an initial solution for propagating it in the 2D optical lattice ($z>0$):
$\phi(\mathbf{x},0)=\phi_{l}^{\mathrm{v}}(\mathbf{x},0)=e^{il\theta}f_{l}^{\mathrm{v}}(r)$.
Thus we solve Eq.(\ref{eq:nonlinear_eq}) for $z>0$, which becomes
a NLSE which the periodic potential $V_{1}(\mathbf{x})$, with the
previous initial condition. This is solved numerically using a standard
split-step Fourier evolution method.

\begin{figure}
\includegraphics[scale=1.1]{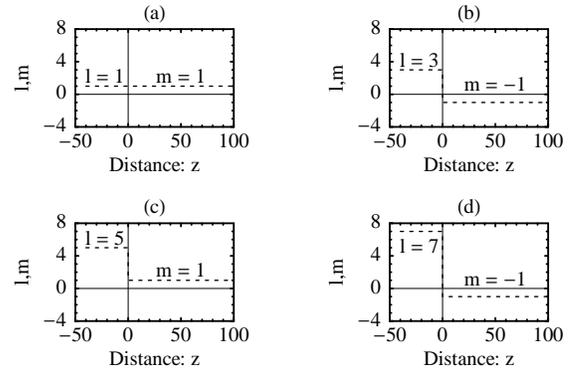}

\caption{Numerical confirmation of the angular momentum {}``pass rule''
for different values of the angular momentum $l$ of the incident
vortex field: (a) $l=1$ and $m=1$; (b) $l=3$ and $m=-1$; (c) $l=5$
and $m=1$; (d) $l=7$ and $m=-1$.\label{cap:confirmation_angular_momentum_rule}}
\end{figure}

According to our previous symmetry arguments, the evolution of the
$\phi$ wave for $z>0$ has to occur in a way that the {}``pass rule''
for angular momentum (\ref{eq:angular_momentum_pass_rule}) is fulfilled.
The $O(2)$ vortex soliton $\phi_{l}^{\mathrm{v}}$ carrying angular momentum
$l$ will excite a propagating wave $\phi_{m}$ for $z>0$ in a representation
of $\mathcal{C}_{4}$ with pseudo-angular momentum $m$ given by Fig.~\ref{cap:allowed_angular_momentum}.
Indeed, numerical evidence of this {}``pass rule'' is obtained by
analyzing the rotational symmetry of the evolving field. By construction,
the input momentum is $l$ since we choose the solution to be of the
form $\phi_{l}^{\mathrm{v}}(\mathbf{x},z)=e^{il\theta}f_{l}^{\mathrm{v}}(r)e^{-i\mu z}$
for $z\leq0$. In order to check the symmetry properties of the solution
for $z>0$, we numerically evaluate the rotated field $\bar{\phi}(r,\theta,z)\equiv\phi(r,\theta+\pi/2,z)$
at every step in $z$ and compare it to its unrotated value $\phi(r,\theta,z)$.
If $\phi$ belongs to the $m$ representation of $\mathcal{C}_{4}$,
$\phi(r,\theta+\pi/2,z)=e^{im\pi/2}\phi(r,\theta,z)$ and the ratio
$\bar{\phi}/\phi$ will have a constant value for all $\mathbf{x}\in\mathbb{R}^{2}$
(with the exception of $\mathbf{x=0}$, where rotations are ill-defined)
and $z>0$: $\bar{\phi}/\phi=e^{im\pi/2}$. If this condition is satisfied
the value of $m$ can be directly extracted from the numerical ratio
$\bar{\phi}/\phi$. Indeed, the independence of the $\bar{\phi}/\phi$ ratio
from transverse coordinates is numerically verified at every axial
step, which permits to evaluate $m$ for different values of $z>0$.
Results are shown in Fig.~\ref{cap:confirmation_angular_momentum_rule}.
These results nicely confirm the general condition (\ref{eq:angular_momentum_pass_rule})
and, more specifically, they satisfy the graphical rule represented
in Fig.~\ref{cap:allowed_angular_momentum} for an $O(2)$-$\mathcal{C}_{4}$
interface.

Once the angular momentum {}``pass rule'' is checked there persist
the question of the fake of the propagating wave in the $\mathcal{C}_{4}$
medium. As predicted by theory, $m$ is numerically conserved during
evolution. However, the asymptotic behavior of the $\phi_{m}$ evolving
field can be very different depending on the parameters of the incident
vortex field (its power $P$ and its propagation constant $\mu$)
and of the characteristics of the periodic potential $V_{1}(\mathbf{x})$
(the potential strength $V_{1}$ and the lattice period $\Lambda$).
Our interest lie in obtaining asymptotic stationary states which can
be described as individual or canonical discrete-symmetry vortices.
This condition implies that the asymptotic field has to present a
single phase singularity. In other words, we want to exclude multi-vortex
or cluster excitations. In order to achieve this feature, we enlarge
the optical lattice (by increasing its period $\Lambda$) according
to the size of the input vortex for increasing values of $l$. Thus,
in our simulations $\Lambda$ is fixed by $l$. 

\begin{figure}
\includegraphics{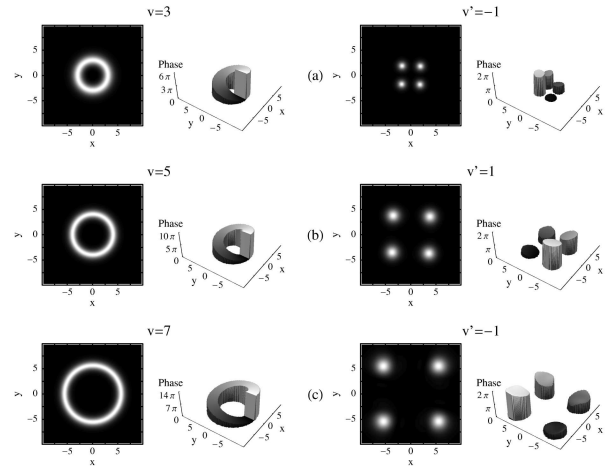}

\caption{Amplitudes and phases of input $O(2)$ vortices and output $\mathcal{C}_{4}$
vortices for different values of the input vorticity $v$: (a) $v=3$
and $v'=-1$ ($P=2.5$, $\mu=2$, $V_{1}=2$, $\Lambda=1.3$); (b)
$v=5$ and $v'=1$ ($P=3.0$, $\mu=3$, $V_{1}=3$, $\Lambda=2.5$);
(c) $v=7$ and $v'=-1$ ($P=3.5$, $\mu=3$, $V_{1}=3$, $\Lambda=3.5$).\label{cap:input_output_vortices}}
\end{figure}

\begin{figure}
\includegraphics{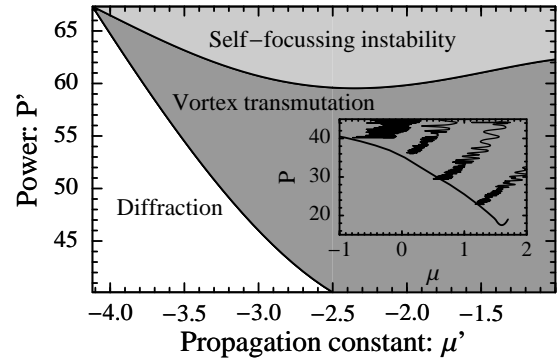}

\caption{{}``Vortex transmutation'' phase diagram for $v=3$ vortices in
an optical lattice with $V_{1}=2$. Inset: Evolution in the $(P',\mu')$
plane of four initial $v=3$ vortices characterized by different initial
$(P,\mu)$ values. The solid line curve corresponds to the stationary
$v'=-1$ vortices as in Ref.\cite{muslimani-josab21_973}.\label{cap:phase_diagram+evolution}}
\end{figure}

After performing many different simulations, we have indeed found
numerical evidence of the {}``vortex transmutation'' phenomenon.
By playing with the input parameters $P$ and $\mu$ and the lattice
strength $V_{1}$ and period $\Lambda$, we have been able to find
asymptotic stationary states $\phi^{\mathrm{v}}_{m}=e^{im\theta}g_{m}(r,\theta)e^{-i\mu'z}$
for different values of the input vorticity value $v=l$. The vorticity
of the output field can be only $v'=\pm1$ because of the vorticity
cutoff for a $\mathcal{C}_{4}$ system (recall that $m=\pm2$ solutions
are not vortices but nodal or dipole-mode solitons \cite{ferrando-arXiv:nlin_0411005}).
In Fig.~\ref{cap:input_output_vortices} we show the amplitudes and
phases of input and output vortices for different input vorticity
values $v$. All of them verify the vorticity {}``pass rule'' (\ref{eq:vorticity_pass_rule}).
The {}``vortex transmutation'' phenomenon only occurs when $|v|>2$.
Similar results are found for the corresponding input anti-vortices
with negative values of $v$. When, for fixed $v=l$ (fixed $\Lambda$),
the election of $P$, $\mu$, and $V_{1}$ is not adequate the asymptotic
solution can be non-stationary. We observe two different scenarios
besides the stationary regime: discrete diffraction of the input wave
in the optical lattice and self-focussing instability leading to filamentation
of the field. A thorough analysis of multiple configurations permits
to elaborate a {}``vortex transmutation'' phase diagram where the
three different regimes can be recognized. As an example, in the phase
diagram shown in Fig.~\ref{cap:phase_diagram+evolution} we observe
the {}``vortex transmutation'' region (shaded) differentiated from
the diffraction (white) and self-focussing instability (light shaded) regions
as a function of the power $P$ and propagation constant $\mu$ of
the input vortex field at fixed $V_{1}$. Analogous phase diagrams
are found for different values of $V_{1}$. It is interesting to analyze
the evolution of different {}``vortex-transmuting'' configurations
by monitoring the evolution of the power $P'(z)$ and average propagation
constant $\mu'(z)\equiv\int\phi^{*}(-i\partial/\partial z)\phi/\int\phi^{*}\phi$
(defined both on the finite domain of the numerical solution) in the
$\mathcal{C}_{4}$ medium. These quantities are $z$-dependent, in
general. However, they become independent of $z$ when we analyze
a stationary solution; thus we expect $(P'(z),\mu'(z))\stackrel{z\rightarrow\infty}{\rightarrow}(P',\mu')$
for asymptotic stationary states. Every input $O(2)$ vortex characterized
by the initial values $(P,\mu)$ defines then a different trajectory
in the $P'$-$\mu'$ plane. In Fig.~\ref{cap:phase_diagram+evolution}
(inset) we show four different trajectories mapping $O(2)$ $v=3$
vortices with different $(P,\mu)$ initial values into asymptotic
$\mathcal{C}_{4}$ vortices with charge $v'=-1$ characterized by
their $(P',\mu')$ values. It can be checked numerically that these
values lie on the same $P'(\mu')$ curve found in Ref.\cite{muslimani-josab21_973}
for stationary vortices with charge $v'=-1$ in
an identical square optical lattice. By launching a whole family of
initial vortices we have been able to asymptotically reproduce the
entire $P'(\mu')$ curve of $\mathcal{C}_{4}$ vortices. It is remarkable
that the asymptotic $\mathcal{C}_{4}$ vortices in the Fig.~\ref{cap:phase_diagram+evolution}
inset have been checked to be stable under small perturbations \cite{muslimani-josab21_973}
whereas the original $O(2)$ ones are not \cite{desyatnikov-arXiv:nlin_0501026}.
Thus the {}``vortex transmutation'' phenomenon not only permits
to change the charge of unstable input vortices but it can also help
to transform them into stable structures. Inversion of the vortex
charge has been observed in {}``noncanonical'' vortices in free-space
and occurs through its dynamic propagation \cite{molina-terriza-prl87_23902}.
Here, however, all vortices involved are {}``canonical'', the key
point to the {}``vortex transmutation'' phenomenon to occur being
the suitable matching between angular and pseudo-angular momentum
at the $O(2)$-$\mathcal{C}_{n}$ interface. Despite our model refers
to an specific system, the theory of transmuting vortices is general
and applies to any system given by an equation of the type $L(|\phi|)\phi=-i\partial\phi/\partial z$
in the presence of an $O(2)$-$\mathcal{C}_{n}$ interface. Hence,
it is expected this phenomenon to occur in a wide variety of physical
systems as those mentioned in the introduction of this Letter.

We are thankful to P. F. de Córdoba for useful discussions and to
grants TIC2002-04527-C02-02 (MCyT and FEDER funds), GV04B-390, and
Grupos03/227 (Generalitat Valenciana). 

\bibliographystyle{apsrev}
\bibliography{bib_general}

\end{document}